\documentclass[12pt,english]{article}
\usepackage[T1]{fontenc}
\usepackage[latin1]{inputenc}
\usepackage{babel}
\usepackage{epsf}

\makeatletter

\providecommand{\LyX}{L\kern-.1667em\lower.25em\hbox{Y}\kern-.125emX\@}
\let\SF@@footnote\footnote
\def\footnote{\ifx\protect\@typeset@protect
    \expandafter\SF@@footnote
  \else
    \expandafter\SF@gobble@opt
  \fi
}
\expandafter\def\csname SF@gobble@opt \endcsname{\@ifnextchar[
  \SF@gobble@twobracket
  \@gobble
}
\edef\SF@gobble@opt{\noexpand\protect
  \expandafter\noexpand\csname SF@gobble@opt \endcsname}
\def\SF@gobble@twobracket[#1]#2{}

\makeatother
\begin{document}

\title{Confined quantum fields under the influence of a uniform magnetic
field}

\author{E. Elizalde\( ^{\dagger } \)\\
 Institut d'Estudis Espacials de Catalunya (IEEC/CSIC) \\
 Edifici Nexus, Gran Capit\`{a} 2-4, 08034 Barcelona \ \& \\
 Departament d'Estructura i Constituents de la Mat\`{e}ria \\
 Facultat de F\'{\i}sica, Universitat de Barcelona \\
 Av. Diagonal 647, 08028 Barcelona, Spain\\
 and \\
 F. C. Santos\( ^{\ddagger } \) and A. C. Tort\( ^{\flat ,\star } \)\\
 Departamento de F\'{\i}sica Te\'{o}rica - Instituto de F\'{\i}sica
\\
 Universidade Federal do Rio de Janeiro\\
 C.P. 68528, 21945-970 Rio de Janeiro, Brazil}

\date{\today{}}

\maketitle
\begin{abstract}
\noindent We investigate the influence of a uniform magnetic field
on the zero-point energy of charged fields of two types, namely, a
massive charged scalar field under Dirichlet boundary conditions and
a massive fermion field under MIT boundary conditions. For the first,
exact results are obtained, in terms of exponentially convergent functions,
and for the second, the limits for small and for large mass are analytically
obtained too. Coincidence with previously known, partial result serves
as a check of the procedure. For the general case in the second situation
---a rather involved one--- a precise numerical analysis is performed. 
\end{abstract}
\vfill

\noindent {\small \( ^{\dagger } \) e-mail: elizalde@ieec.fcr.es}
\\
 {\small \( ^{\ddagger } \) e-mail: filadelf@if.ufrj.br} \\
 {\small \( ^{\flat } \) e-mail: tort@if.ufrj.br} \\
 {\small \( ^{\star } \) Present address: Institut d'Estudis Espacials
de Catalunya (IEEC/CSIC) Edifici Nexus, Gran Capit\`{a} 2-4, 08034
Barcelona, Spain. E-mail: visit11@ieec.fcr.es\par{}}{\small \par}

\vfill

\newpage

\section{Introduction}

The Casimir effect \cite{HBGCasimir48} is a very fundamental feature
common to all quantum field theories, which arises in particular,
as is well known, when there is a departure from the topology of the
ordinary flat spacetime towards non-trivial topologies. It has been
studied intensively in the last years due to its importance in elementary
particle physics, cosmology and condensed matter physics ---see Ref.
\cite{BMM2001} for a recent review on the theoretical and experimental
aspects of this effect, and also Refs. \cite{OlderRBooks}.

Whenever we deal with a confined charged quantum field, it is a natural
question to ask for the influence of external fields on their zero-point
oscillations and to investigate its consequences on the Casimir effect.
The influence of external fields on zero-point oscillations of unconfined
charged bosonic and fermionic fields and the construction of the corresponding
effective field theories (at the one-loop level) is an honorable subject
and it is linked to the pioneering works of the thirties by Heisenberg,
Kockel and Euler, Heisenberg and Euler, and Weisskopf \cite{Pioneers}
and of the fifties by Schwinger \cite{Schwinger1951}. More recently,
the influence of a uniform magnetic field was extensively treated
by Dittrich and Reuter in the monograph \cite{RD}, and zeta function
techniques were employed to construct effective Lagrangians for scalars
and Dirac fields for \( d=2,3 \) and higher dimensions in constant
background fields (see Refs. \cite{Elisalde94} and references therein).
Ambjorn and Wolfram considered the influence of an electrical field
on the vacuum fluctuations of a charged scalar field \cite{AW}. Outside
the context of pure QED, the influence on the vacuum energy density
of a real scalar field due to an arbitrary background of an also real
scalar field interpreted as a substitute for hard boundary conditions
was considered in Refs. \cite{Bordag1995,BordagLindig96,ActorBender95}.
In the gravitational case, Elizalde and Romeo investigated the issue
of a neutral scalar field in the presence of a static external gravitational
field \cite{ER1997}. A general and valuable discussion of these aspects
of quantum field theory can be found in Ref. \cite{Gribetal} too.

The influence of external fields on zero-point oscillations of quantum
fields confined by hard boundary conditions, however, has been investigated
in some particular cases only. For example, the influence of a uniform
magnetic field on the Casimir effect was investigated in Ref. \cite{CPCFACT2001}
in the cases of a massive fermion field and of a charged scalar field
both submitted to anti-periodic boundary conditions, and in Ref. \cite{CPCFMRNACT1999}
in the case of a massive charged scalar field submitted to Dirichlet
boundary conditions. In the cases considered by those authors it was
formally shown that the fermionic Casimir effect is enhanced by the
applied magnetic field while the bosonic one is inhibited. In Refs.
\cite{CPCFACT2001,CPCFMRNACT1999}, Schwinger's proper-time method
\cite{Schwinger1951} was employed and explicit analytical results
obtained for particular regimes of a conveniently defined dimensionless
parameter \( \mu  \) and magnitude of the external field. In the
weak magnetic field regime, some aspects of the material properties
of confined charged fields were investigated in Refs. \cite{MatProp}.

In this work we wish to resume the investigation of the influence
of an external magnetic field on the zero-point oscillations of charged
confined quantum fields and consider the case of a massive fermion
field under a uniform magnetic field and constrained by boundary conditions
of the MIT type, which states that the fermionic current through an
hypothetical confining surface must be zero \cite{Johnson75}. This
is not an academic question since, for instance, in the bag model
of hadrons \cite{Johnson75,BagModel}, we can expect the zero-point
oscillations of the quark fields to be influenced by the strong electric
and magnetic fields that permeate the interior of the hadronic bag
(see for example \cite{Kabatetal} and references therein). We wish
also to reconsider the case of a massive charged scalar field under
a uniform magnetic field by investigating regimes not considered in
Ref. \cite{CPCFMRNACT1999}. In order to evaluate the relevant Casimir
energies it is convenient, especially in the case of the massive fermion
field with MIT boundary conditions, to follow the spirit of the representation
of spectral sums as contour integrals \cite{relatedpapers}, which
allows for the incorporation of the boundary conditions in a smooth
way. Here we will employ a relatively simple variant of this general
procedure. This variant is described in \cite{ST2002} and it will
be here applied without further explanation to the cases at hand (the
interested reader should consult that paper for details). The outline
of the present work is as follows. In Section 2 we apply our calculational
tools to the charged scalar field case. In Section 3, we consider
the case of the charged fermion field. For both cases, detailed numerical
analysis and some special analytical limits are provided, together
with an explanation of the results obtained and specific comparisons
with other results. Section 4 is devoted to final remarks. Throughout
the paper we employ natural units (\( \hbar =c=1.) \)

\section{The vacuum energy of a charged scalar field with Dirichlet boundary
conditions and under the influence of a uniform magnetic field}

Let us first briefly consider a charged scalar field under Dirichlet
boundary conditions imposed on the field on two parallel planes separated
by the distance \( \ell  \) whose side \( L \) satisfies the condition
\( L\gg \ell  \). Suppose initially that there is no external field.
The unregularised Casimir energy is given by \cite{ST2002}%
\footnote{Notice that in the absence of external fields we could have started
from Eq. (21) in Ref. \cite{ST2002}, but for our purposes here it
is more convenient to start from Eq. (18).
}\begin{equation}
\label{startingpoint}
E_{0}\left( \ell ,\mu \right) =\alpha \frac{L^{2}}{\left( 2\pi \right) ^{3}}\int d^{3}p\, \log \left[ 1+\frac{K_{1}\left( z\right) }{K_{2}\left( z\right) }\right] ,
\end{equation}
 where \( K_{1}\left( z\right)  \) and \( K_{2}\left( z\right)  \)
are functions constructed from the boundary conditions as described
in Ref. \cite{ST2002}. The dimensionless parameter \( \alpha  \)
takes into account the internal degrees of freedom of the quantum
field. For Dirichlet boundary conditions we begin by writing \begin{equation}
F\left( z\right) =\sin \, z.
\end{equation}
 Since \( z=0 \) is a root of \( F \)\( \left( z\right)  \) we
divide this function by \( z \), thus removing \( z=0 \) from the
set of roots without introducing a new singularity: This is equivalent
to removing the zero mode. Define\begin{equation}
G\left( z\right) =\frac{\sin \, z}{z}.
\end{equation}
 The function \( K\left( z\right) =K_{1}\left( z\right) +K_{2}\left( z\right)  \),
with \( K_{1}\left( z\right) =K_{2}\left( -z\right)  \), is then
obtained by performing the substitution \( z\rightarrow iz \), that
is \begin{equation}
\label{Kboson}
K\left( z\right) :=G\left( iz\right) =\frac{e^{z}-e^{-z}}{2z},
\end{equation}
 where \( z \) is the function\begin{equation}
z=z\left( p_{1,}p_{2},p_{3}\right) =\ell \sqrt{p_{1}^{2}+p_{2}^{2}+p_{3}^{2}+m^{2}}.
\end{equation}
 Since \( K_{1}\left( z\right) =-e^{-z}/2z \) and \( K_{2}\left( z\right) =e^{z}/2z \)
we have\begin{equation}
E_{0}\left( \ell ,\mu \right) =\alpha \frac{L^{2}}{2}\int \frac{d^{3}p}{\left( 2\pi \right) ^{3}}\, \log \left[ 1-e^{-2z}\right] .
\end{equation}
 To shorten these initial steps let us set \( m=0 \). Then after
expanding the log we obtain, simply, \begin{equation}
E_{0}\left( \ell \right) =-\frac{\alpha L^{2}}{4\pi ^{2}\ell ^{3}}\sum _{k=1}^{\infty }\frac{1}{k}\int _{0}^{\infty }dx\, x^{2}e^{-2kx},
\end{equation}
 where we have defined \( x:=p_{3}\ell  \). The integral can be evaluated
with the help of the Mellin transform\begin{equation}
A^{-s}\Gamma \left( s\right) =\int dt\, t^{s-1}e^{-At}.
\end{equation}
 and after, simple manipulations, we obtain the well known result\begin{equation}
E_{0}\left( \ell \right) =-\frac{\alpha \, L^{2}\, \pi ^{2}}{1440\, \ell ^{3}}.
\end{equation}
 For \( \alpha =2 \) we get the result valid for photons. A more
complex example of the usefulness of Eq. (\ref{startingpoint}) is
provided in Ref. \cite{ST2002}.

Let us consider now the same type of field and boundary conditions
in the presence of an external magnetic field which we will suppose
to be uniform and perpendicular to the two Dirichlet planes. We will
assume also that \( eB \) points towards the positive \( OX_{3} \)
direction. Equation (\ref{startingpoint}) reads now \begin{equation}
E_{0}\left( \ell ,\mu ,eB\right) =\alpha \left( \frac{eB}{2\pi }\right) \frac{L^{2}}{2}\sum _{n=0}^{\infty }\int _{-\infty }^{\infty }\frac{dp_{3}}{2\pi }\log \left[ 1-e^{-2z}\right] ,
\end{equation}
 where we have taken into account that for a charged spin zero boson
the Landau levels are given by \begin{equation}
p_{1}^{2}+p_{2}^{2}=eB\left( 2n+1\right) ,\, \, \, \, \, \, \, \, n=0,1,2,3,...\, .
\end{equation}
 The factor \( eB/2\pi  \) is the degeneracy factor and for a charged
scalar field \( \alpha =2 \). Hence, the function \( z \) reads
now \begin{equation}
z=z\left( p_3,n\right) :=\sqrt{\ell ^{2}p_3^{2}+eB\ell ^{2}\left( 2n+1\right) +\mu ^{2}},
\end{equation}
 where \( \mu :=\ell m \). It is convenient to define\begin{equation}
M_{n}:=\sqrt{\left( 2n+1\right) eB\ell ^{2}+\mu ^{2}}
\end{equation}
 and write\begin{equation}
E_{0}\left( \ell ,\mu ,eB\right) =\frac{eBL^{2}}{2\pi ^{2}\ell }\sum _{n=0}^{\infty }I_{n}\left( M_{n}\right) ,
\end{equation}
 where\begin{equation}
I_{n}\left( M_{n}\right) =\int _{0}^{\infty }dx\, \log \left[ 1-e^{-2\sqrt{x^{2}+M_{n}^{2}}}\right] ,
\end{equation}
 and we have set \( x:=p_{3}\ell  \). Expanding the log and introducing
the variable \( \omega  \), defined by \( \omega :=\sqrt{x^{2}+M_{n}^{2}} \),
we end up with\begin{equation}
E_{0}\left( \ell ,\mu ,eB\right) =\frac{eBL^{2}}{2\pi ^{2}\ell }\sum _{n=0}^{\infty }\sum _{k=1}^{\infty }\frac{1}{k}I_{kn}\left( M_{n}\right) ,
\end{equation}
 where\begin{equation}
I_{kn}\left( M_{n}\right) :=\int _{M_{n}}^{\infty }d\omega \, \omega \left( \omega +M_{n}\right) ^{-1/2}\left( \omega -M_{n}\right) ^{-1/2}e^{-2k\omega }.
\end{equation}
 In order to evaluate this integral, we first introduce an auxiliary
integral defined by \begin{equation}
I_{kn}\left( M_{n},\lambda \right) :=\int _{M_{n}}^{\infty }d\omega \, \left( \omega +M_{n}\right) ^{-1/2}\left( \omega -M_{n}\right) ^{-1/2}e^{-2k\omega \lambda }.
\end{equation}
 To evaluate the auxiliary integral we make use of (\emph{c.f.} formula
3.384.3 in \cite{Grad94})\begin{eqnarray}
\int _{\mu _{1}}^{\infty }dx\, \left( x+\beta \right) ^{2\nu -1}\left( x-\mu _{1}\right) ^{2\rho -1}e^{-\mu _{2}x}=\frac{\left( \mu _{1}+\beta \right) ^{\nu +\rho -1}}{\mu _{2}^{\nu +\rho }}\exp \left[ \frac{\left( \beta -\mu _{1}\right) }{2}\mu _{2}\right]  &  & \nonumber \\
\times \Gamma \left( 2\rho \right) W_{\nu -\rho ,\nu +\rho -\frac{1}{2}}\left( \mu _{1}\mu _{2}+\beta \mu _{2}\right) , &  & \label{Grad01} 
\end{eqnarray}
 which holds for \( \mu _{1}>0 \), \( \left| \mbox {Arg}\, \left( \beta +\mu _{1}\right) \right| <\pi , \)
Re \( \mu _{2}>0 \) and Re \( \rho >0 \). The auxiliary integral
then reads \begin{equation}
\label{IKlambda}
I_{kn}\left( M_{n},\lambda \right) =\frac{\left( 2\mu \right) ^{-1/2}}{\left( 2k\lambda \right) ^{1/2}}\Gamma \left( 1/2\right) W_{0,0}\left( 4kM_{n}\lambda \right) ,
\end{equation}
 where \( W_{\mu ,\lambda }\left( z\right)  \) is the Whittaker function
\cite{Whittaker}. The Whittaker \( W_{0,0}\left( z\right)  \) function
is related to the modified Bessel function of the third kind through
\cite{Grad94}\begin{equation}
W_{0,0}\left( 4kM_{n}\lambda \right) =\frac{\left( 4kM_{n}\lambda \right) ^{1/2}}{\pi ^{1/2}}K_{0}\left( 2kM_{n}\lambda \right) .
\end{equation}
 Taking the derivative of Eq. (\ref{IKlambda}) with respect to \( \lambda  \),
and setting \( \lambda =1 \) to obtain \( I_{kn}\left( M_{n}\right)  \),
we have\begin{equation}
I_{kn}\left( M_{n}\right) =2M_{n}\frac{d}{dz}K_{0}\left( z=2kM_{n}\lambda \right) \left| _{\lambda =1}\right. =-M_{n}K_{1}\left( 2kM_{n}\right) .
\end{equation}
 Hence the vacuum energy reads \begin{equation}
\label{ScalarBEnergy0}
E_{0}\left( \ell ,\mu ,eB\right) =-\frac{eBL^{2}}{2\pi ^{2}\ell }\sum _{n=0}^{\infty }M_{n}\, \sum _{k=1}^{\infty }\frac{1}{k}K_{1}\left( 2kM_{n}\right) ,
\end{equation}
 or, more explicitly, \begin{eqnarray}
 &  & E_{0}\left( \ell ,\mu ,eB\right) =-\frac{eBL^{2}}{2\pi ^{2}\ell }\sqrt{eB\ell ^{2}+\mu ^{2}}\, \sum _{k=1}^{\infty }\frac{1}{k}K_{1}\left( 2k\sqrt{eB\ell ^{2}+\mu ^{2}}\right) \nonumber \\
 & - & \frac{eBL^{2}}{2\pi ^{2}\ell }\, \sum _{n=1}^{\infty }\sqrt{\left( 2n+1\right) eB\ell ^{2}+\mu ^{2}}\, \sum _{k=1}^{\infty }\frac{1}{k}K_{1}\left( 2k\sqrt{\left( 2n+1\right) eB\ell ^{2}+\mu ^{2}}\right) \label{ScalarBEnergy} 
\end{eqnarray}
 This representation is an alternative to the one obtained in \cite{CPCFMRNACT1999}.
In a very strong field, such that \( eB\ell ^{2}\gg \mu ^{2} \),
the decaying exponential behavior of the Bessel functions of the third
kind allows us to keep only the term corresponding to \( k=1 \) in
the first parcel of the vacuum energy and, thus, in this limit we
have\begin{equation}
\label{ScalarBenergylm}
\frac{E_{0}\left( \ell ,eB\right) }{L^{2}}\approx -\frac{\left( eB\ell ^{2}\right) ^{5/4}}{\pi ^{3/2}\ell ^{3}}e^{-2\sqrt{eB\ell ^{2}}},
\end{equation}
 in agreement with \cite{CPCFMRNACT1999}. Notice that for zero magnetic
field the vacuum energy as given by Eqs. (\ref{ScalarBEnergy}) and
(\ref{ScalarBenergylm}) is zero, that is, the zero of the energy
is automatically shifted with respect to the vacuum energy in the
absence of the external field (which serves, therefore, as the natural
origin of energies).

\subsection{Numerical analysis for the scalar case}

For arbitrary values of the parameter \( \mu  \) and of the scaled
field \( eB\ell ^{2} \), a numerical evaluation of Eq. (\ref{ScalarBEnergy0})
is still possible. The graphs of Figs. 1a, 1b, 2, and 3 exhibit some
representative examples.

\begin{figure}[htb]
\centerline{\epsfxsize=8cm \epsfbox{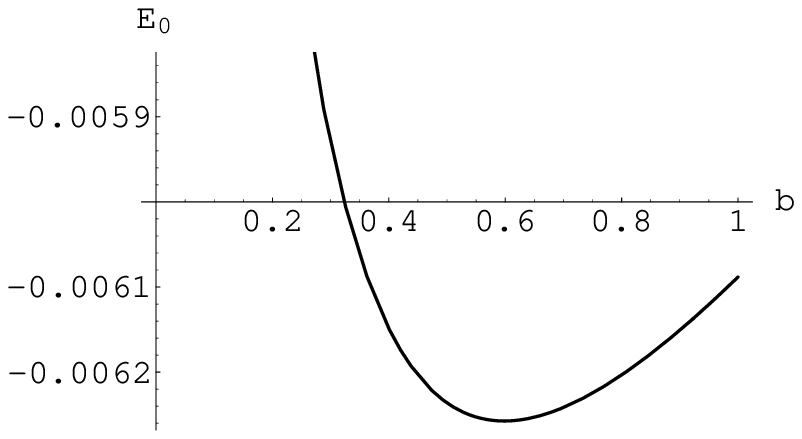} \hspace{8mm} \epsfxsize=8cm
\epsfbox{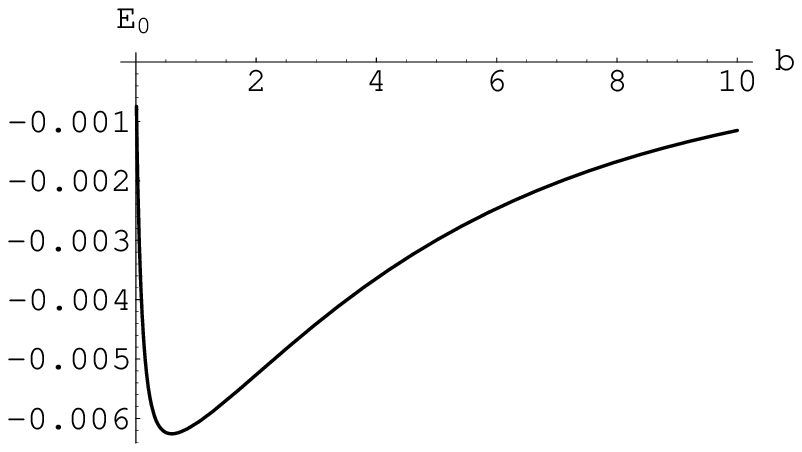}}
\caption{{\protect\small Plot of the energy in terms of the dimensionless
quantity $b (=eB\ell ^{2})$ corresponding to the magnetic field,
for a fixed value
of the dimensionless mass $\mu$ (here $\mu =1$). Fig. 1a shows in detail
the formation of a smooth minimum and its precise value.}} \label{f1}
\end{figure}
\begin{figure}[hbt]
\centerline{\epsfxsize=10cm \epsfbox{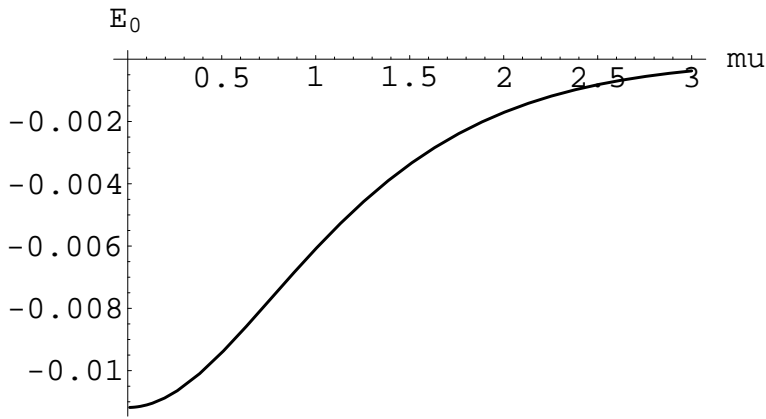}}
\caption{{\protect\small Plot of the energy in terms of the dimensionless
mass $\mu$, for a fixed value
of the dimensionless magnetic field $b$ (here $b =1$). For a wide range of
values of $b$ one gets a curve with a similar shape.}} \label{f2}
\end{figure}
\begin{figure}[htb]
\centerline{\epsfxsize=10cm \epsfbox{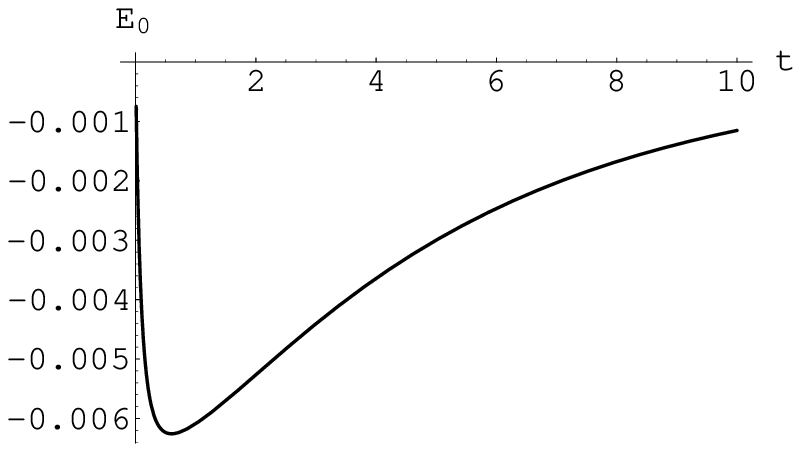}}
 \caption{{\protect\small Plot of the energy in terms of the dimensionless
variable
$t=b/\mu^2$, that measures the magnitude of the dimensionless
magnetic field $b$ in terms of the dimensionless mass $\mu^2$.
It is clear that, at $\mu =1$ we recover the same shape as in
Fig. 1.}} \label{f3}
\end{figure}
 In the first two graphs (Fig. 1) we plot a dimensionless version
of Eq. (\ref{ScalarBEnergy0}) as a function of the scaled magnetic
field \( b:=eB\ell ^{2} \), for \( \mu =1 \) and a convenient range
of \( b \) that clearly shows the behavior of the energy (Fig. 1b).
Notice that initially the external field decreases the value of the
vacuum energy up to a certain value of \( b \) for which the vacuum
energy attains a minimum (which is most clearly depicted in Fig. 1a).
After reaching this point, the energy grows as remarked in Ref. \cite{CPCFMRNACT1999}.
This situation is the opposite of what happens with a fermionic field,
in which case the energy decreases linearly with \( b \), for large
values of \( b \) (as will be clearly seen in the next section).

Fig. 2 shows the behavior of the energy in terms of the mass \( \mu  \)
for a fixed value of the magnetic field \( b \) (chosen here as \( b=1 \),
but the shape of the curve is very similar for a wide range of values
of \( b \)). It starts at a non-zero value, for zero magnetic field.
Also this is quite different from the behavior in the case of a fermionic
field.

Finally, in Fig. 3 the energy is depicted vs the variable \( t=b/\mu ^{2} \),
that measures the magnitude of the dimensionless magnetic field \( b \)
in terms of the dimensionless mass \( \mu ^{2} \). Notice that, although
for \( \mu =1 \) we obtain the same curve as in Fig. 1, for any other
value of \( \mu  \) the precise shape of Fig. 3 is valuable, as a
way of representing the energy in terms of the unique quantity \( t \)
(the magnetic field in units of mass squared).

Families of curves corresponding to different values of one of the
variables, and also two-dimensional graphs are easy to obtain, in
a reasonable amount of time, from our formulas in this section. \vspace{0.3cm}

\section{Confined fermion field in a uniform magnetic field}

As in the case of the charged bosons, the uniform magnetic field is
here perpendicular to the (hypothetical) parallel MIT constraining
surfaces. The distance between them is \( \ell  \) and \( eB \)
is again supposed to point towards the positive \( OX_{3} \) direction.
The starting expression reads\begin{equation}
E_{0}\left( \ell ,\mu ,eB\right) =-2\times \frac{1}{2}\times \left( \frac{eBL^{2}}{2\pi }\right) \sum _{n=0}^{\infty }\, \sum _{\alpha \in \left\{ -1,1\right\} }\, I_{n\alpha },
\end{equation}
 where the multiplicative factor 2 takes into account the particle
and antiparticle states, and we have defined\begin{equation}
I_{n\alpha }:=\int _{-\infty }^{\infty }\frac{dp_{3}}{2\pi }\log \left[ 1-\frac{z-\mu }{z+\mu }e^{-2z}\right] ,
\end{equation}
 where again (with \( \mu :=\ell m \)) \begin{equation}
z=z\left( q,n,\alpha \right) :=\sqrt{\ell ^{2}p_{3}^{2}+\left( 2n+1-\alpha \right) eB\ell ^{2}+\mu ^{2}}\, \, \, \, \, \, \, n=0,1,2,3...\, .
\end{equation}
 Expanding the log as before we obtain\begin{eqnarray}
\log \left[ 1-\frac{z-\mu }{z+\mu }e^{-2z}\right] =\sum ^{\infty }_{k=1}\frac{\left( -1\right) ^{k+1}}{k}\left[ z+\mu \right] ^{-k} &  & \nonumber \\
\times \left[ z-\mu \right] ^{k}\, e^{-2kz}, &  & 
\end{eqnarray}
 and performing the sum over \( \alpha  \), we arrive at\begin{equation}
\label{EOprime}
E_{0}\left( \ell ,\mu ,eB\right) =-2\times \frac{eBL^{2}}{2\pi ^{2}\ell }\sum ^{\infty \, \, \, \, \, \, '}_{p=-1}\, \sum _{k=1}^{\infty }\, \frac{\left( -1\right) ^{k+1}}{k}I_{pk}\left( M_{p}\right) ,
\end{equation}
 where\begin{equation}
I_{pk}\left( M_{p}\right) :=\int _{0}^{\infty }\, dx\, \left[ \left( x^{2}+M_{p}^{2}\right) ^{1/2}+\mu \right] ^{-k}\left[ \left( x^{2}+M_{p}^{2}\right) ^{1/2}-\mu \right] ^{k}e^{-2k\left( x^{2}+M_{p}^{2}\right) ^{1/2},}
\end{equation}
 with \( x:=p_{3}\ell  \) and\begin{equation}
M_{p}^{2}:=2\left( p+1\right) eB\ell ^{2}+\mu ^{2},\, \, \, \, \, p=-1,0,1,2,3,...\, .
\end{equation}
 The prime in Eq. (\ref{EOprime}) means that the term corresponding
to \( p=-1 \) must be multiplied by the factor \( 1/2 \). Let us
define, as before, a new variable \( \omega  \) through \( \omega \: =\left( x^{2}+M_{p}^{2}\right) ^{1/2} \).
Then\begin{equation}
\label{thehardone}
I_{pk}\left( M_{p}\right) =\int _{M_{p}}^{\infty }\, d\omega \, \omega \left( \omega +M_{p}\right) ^{-1/2}\left( \omega -M_{p}\right) ^{-1/2}\left( \omega +\mu \right) ^{-k}\left( \omega -\mu \right) ^{k}e^{-2k\omega }.
\end{equation}
 The integral defined by Eq. (\ref{thehardone}) is non-trivial. Here
we will solve it analytically in two limits, to wit: the large-- and
the small--\( \mu  \) limits. A numerical integration of Eq. (\ref{thehardone}),
however, is feasible to conveniently complete the analysis. We will
show the results later.

\subsection{The limit \protect\protect\protect\( \mu \ll 1\protect \protect \protect \)}

If we set \( \mu \approx 0 \) we have to solve a much simpler integral
for \( I_{pk}\left( M_{p}\right)  \). In fact,\begin{equation}
I_{pk}\left( M_{p}\right) =\int _{M_{p}}^{\infty }\, d\omega \, \frac{\omega e^{-2k\omega }}{\sqrt{\omega ^{2}-M_{p}^{2}}}.
\end{equation}
 This integral can be evaluated with the help of (\emph{cf} formula
3.365.2 of \cite{Grad94})\begin{equation}
\int _{a}^{\infty }\frac{x\, e^{-bx}}{\sqrt{x^{2}-a^{2}}}=aK_{1}\left( ab\right) ,\, \, \, \, \, \, a>0,\, \, Re\, a>0.
\end{equation}
 For \( a=0 \) this integral reads \begin{equation}
\int _{0}^{\infty }e^{-bx}=\frac{1}{b},
\end{equation}
 as can be easily seen by recalling that, for \( z\rightarrow 0 \),
\( K_{1}\left( z\right) \approx 1/z \). There are three types of
contributions to the vacuum energy: corresponding to \( p=-1 \),
\( p=0 \), and \( p=1,2,3,... \), respectively. They read\begin{eqnarray}
E_{0}\left( \ell ,\mu \approx 0,eB\right) =-\frac{eBL^{2}}{2\pi ^{2}\ell }\, \sum _{k=1}^{\infty }\, \frac{\left( -1\right) ^{k+1}}{k}I_{-1k}-\frac{eBL^{2}}{\pi ^{2}\ell }\, \sum _{k=1}^{\infty }\, \frac{\left( -1\right) ^{k+1}}{k}I_{0k} &  & \nonumber \\
-\frac{eBL^{2}}{\pi ^{2}\ell }\, \sum _{p=1}^{\infty }\, \sum _{k=1}^{\infty }\, \frac{\left( -1\right) ^{k+1}}{k}I_{pk} & , & 
\end{eqnarray}
 where the integrals in the partial sums are given by \begin{equation}
I_{-1k}=\frac{1}{2k}\, ,
\end{equation}
\begin{equation}
I_{0k}=\sqrt{2eB\ell ^{2}}\, K_{1}\left( 2k\sqrt{2eB\ell ^{2}}\right) \, ,
\end{equation}
\begin{equation}
I_{pk}=\sqrt{2\left( p+1\right) eB\ell ^{2}}\, K_{1}\left( 2k\sqrt{2\left( p+1\right) eB\ell ^{2}}\right) \, .
\end{equation}
 It follows then that the first sum can be exactly evaluated by using
the eta function, \( \eta _{R}\left( z\right) =\left( 1-2^{1-s}\right) \zeta _{R}\left( s\right)  \).
In fact, we have\begin{equation}
\sum _{k=1}^{\infty }\, \frac{\left( -1\right) ^{k+1}}{k}I_{-1k}=\frac{\pi ^{2}}{24}.
\end{equation}
 Therefore, the Casimir energy in this limit reads\begin{eqnarray}
E_{0}\left( \ell ,\mu \approx 0,eB\right) =-\frac{eBL^{2}}{48\, \ell }-\frac{eBL^{2}}{\pi ^{2}\ell }\, \sum _{k=1}^{\infty }\, \frac{\left( -1\right) ^{k+1}}{k}\sqrt{2eB\ell ^{2}}\, K_{1}\left( 2k\sqrt{2eB\ell ^{2}}\right)  &  & \nonumber \\
-\frac{eBL^{2}}{\pi ^{2}\ell }\, \sum _{p=1}^{\infty }\, \sum _{k=1}^{\infty }\, \frac{\left( -1\right) ^{k+1}}{k}\sqrt{2\left( p+1\right) eB\ell ^{2}}\, K_{1}\left( 2k\sqrt{2\left( p+1\right) eB\ell ^{2}}\right) . &  & \label{Ferenergymz} 
\end{eqnarray}
 From this result we can easily extract the limit \( eB\ell ^{2}\gg 1 \).
Due to the decaying exponential behavior of the modified Bessel function,
the leading contribution is given by the first term in Eq. (\ref{Ferenergymz}).
Hence\begin{equation}
\label{VacEBmz}
E_{0}\left( \ell ,\mu \approx 0,eB\right) \approx -\frac{eBA}{48\, \ell },\, \, \, \, \, \, \, \, \, \, \, eB\ell ^{2}\gg 1.
\end{equation}
 This result has the same sign ---and one fourth of the magnitude---
of the corresponding problem with anti-periodic boundary conditions
in this same limit \cite{CPCFACT2001}. This difference can be readily
understood if we recall that the anti-periodicity length corresponds
to \( 2\ell  \), and that the number of allowed modes is twice the
number of modes of the MIT case for massless fermions. Therefore,
the result given by Eq. (\ref{VacEBmz}), in that particular limit,
is compatible with the one given in Ref. \cite{CPCFACT2001}.

\subsection{The limit \protect\protect\protect\( \mu \gg 1\protect \protect \protect \)}

In this limit only the term corresponding to \( p=-1 \), \( M_{-1}=\mu  \),
contributes. Therefore, we have to calculate \begin{equation}
I_{-1k}\left( \mu \right) =\int _{\mu }^{\infty }\, d\omega \, \omega \left( \omega +\mu \right) ^{-k-1/2}\left( \omega -\mu \right) ^{k-1/2}e^{-2k\omega }
\end{equation}
 The trick to evaluate this integral we have learned it above. One
should write the auxiliary integral\begin{equation}
I_{-1k}\left( \mu ,\lambda \right) :=\int _{\mu }^{\infty }\, d\omega \, \left( \omega +\mu \right) ^{-k-1/2}\left( \omega -\mu \right) ^{k-1/2}e^{-2k\omega \lambda }
\end{equation}
 and evaluate it with the help of formula given by Eq. (\ref{Grad01}).
The result is\begin{equation}
I_{-1k}\left( \mu ,\lambda \right) =\frac{\Gamma \left( k+1/2\right) }{\left( 4\mu k\lambda \right) ^{1/2}}W_{-k0}\left( 4\mu k\lambda \right) .
\end{equation}
 For large values of the argument, the Whittaker function behaves
as \begin{equation}
W_{\nu \mu }\left( z\right) \approx z^{\nu }e^{-z/2},
\end{equation}
 hence for \( \mu \gg 1 \) we can write\begin{equation}
I_{-1k}\left( \mu ,\lambda \right) \approx \Gamma \left( k+1/2\right) \frac{e^{-2\mu k\lambda }}{\left( 4\mu k\lambda \right) ^{k+1/2}},
\end{equation}
 and since \begin{equation}
I_{-1k}\left( \mu ,\lambda \right) =-\frac{1}{2k}\frac{d}{d\lambda }\left[ \Gamma \left( k+1/2\right) \frac{e^{-2\mu k\lambda }}{\left( 4\mu k\lambda \right) ^{k+1/2}}\right] _{\lambda =1},
\end{equation}
 we finally have \begin{equation}
I_{-1k}\left( \mu ,\lambda \right) =\frac{\Gamma \left( k+1/2\right) }{2^{2k+2}\, k^{k+3/2}}\left[ \frac{\left( k+1/2\right) }{\mu ^{k+1/2}}+\frac{2k}{\mu ^{k-1/2}}\right] .e^{-2\mu k}
\end{equation}
 The vacuum energy in this regime is given bay \begin{equation}
E_{0}\left( \ell ,\mu \gg 1,eB\right) \approx -\frac{eBL^{2}}{8\pi ^{2}\ell }\, \sum _{k=1}^{\infty }\, \frac{\left( -1\right) ^{k+1}}{2^{2k}k^{k*5/2}}\Gamma \left( k+1/2\right) \left[ \frac{\left( k+1/2\right) }{\mu ^{k+1/2}}+\frac{2k}{\mu ^{k-1/2}}\right] e^{-2\mu k},
\end{equation}
 a result that is new and incorporates the first mass corrections.

The most relevant one is the \( k=1 \) term ,which yields \begin{equation}
E_{0}\left( \ell ,\mu \gg 1,eB\right) \approx -\frac{eBL^{2}}{32\pi ^{3/2}\ell }\frac{e^{-2\mu }}{\mu ^{1/2}},
\end{equation}
 that is sufficiently small. This is not the same result as the one
obtained for anti-periodic boundary conditions though the damping
exponential appears in both cases. The factors multiplying the exponential
are different, it must be remembered though that for \( \mu \neq 0 \)
the MIT and the AP spectra are not comparable. When \( \mu =0 \)
the MIT and the AP spectra for each component of the fermion field
differ by a factor of 4.

\subsection{Numerical analysis for arbitrary mass}

For arbitrary values of the mass and the magnetic field or, correspondingly,
of their dimensionless counterparts, \( \mu  \) and \( eB\ell ^{2} \),
the integral defined by (\ref{thehardone}) is very hard to evaluate
analytically. In fact, even if we would give it in terms of hypergeometric
functions this would not improve our knowledge of the dependence of
the energy on the mass and the magnetic field. Fortunately, a numerical
analysis based on Eq. (\ref{EOprime}) is possible and leads to very
precise, easily understandable results.

To this end, first we rewrite Eq. (\ref{EOprime}) under the form
\begin{equation}
\frac{\ell ^{3}E_{0}\left( t,\mu \right) }{L^{2}}=-\frac{t\mu ^{3}}{\pi ^{2}}\sum ^{\infty \, \, \, \, \, \, '}_{p=-1}\, \sum _{k=1}^{\infty }\, \frac{\left( -1\right) ^{k+1}}{k}J_{pk}\left( t,\mu \right) ,
\end{equation}
 where \begin{eqnarray}
J_{pk}\left( t,\mu \right) :=\int _{0}^{\infty }\, dy\, \left[ \left( y^{2}+2\left( p+1\right) t+1\right) ^{1/2}+1\right] ^{-k}\left[ \left( y^{2}+2\left( p+1\right) t+1\right) ^{1/2}-1\right] ^{k} & \nonumber \\
\times e^{-2k\mu \left( y^{2}+2\left( p+1\right) t+1\right) ^{1/2}} & \label{NumAnalF} 
\end{eqnarray}
 with \( y:=x/\mu  \) and \( t=eB\ell ^{2}/\mu ^{2} \).
 
\begin{figure}[htb]
\centerline{\epsfxsize=8cm \epsfbox{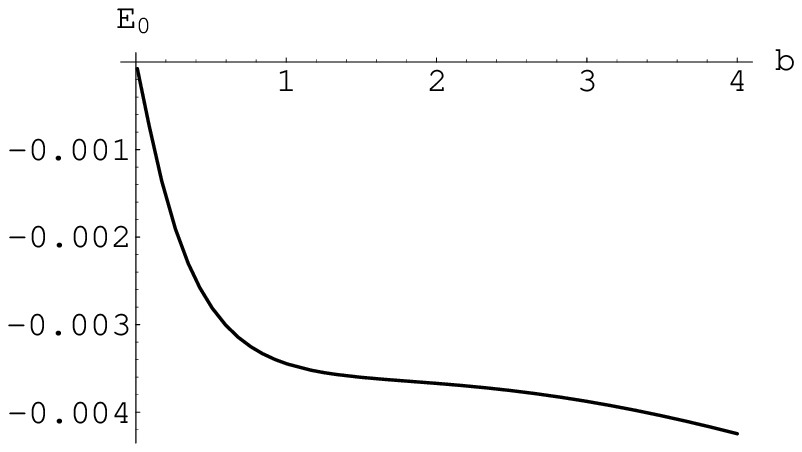} \hspace{8mm} \epsfxsize=8cm
\epsfbox{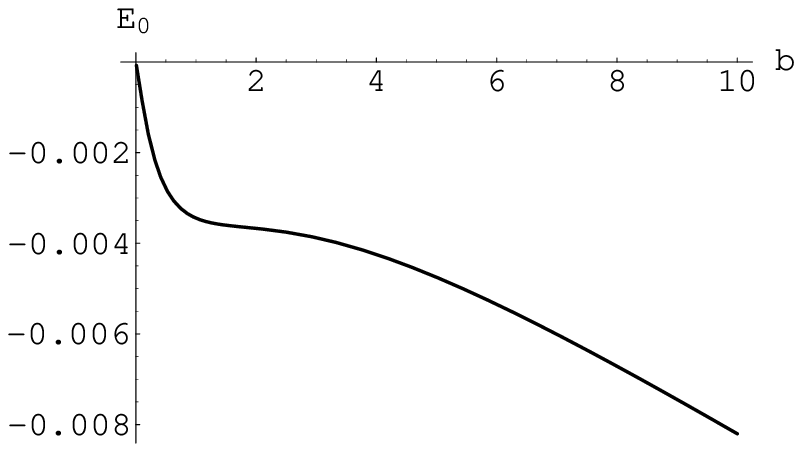}}
\centerline{\epsfxsize=9cm \epsfbox{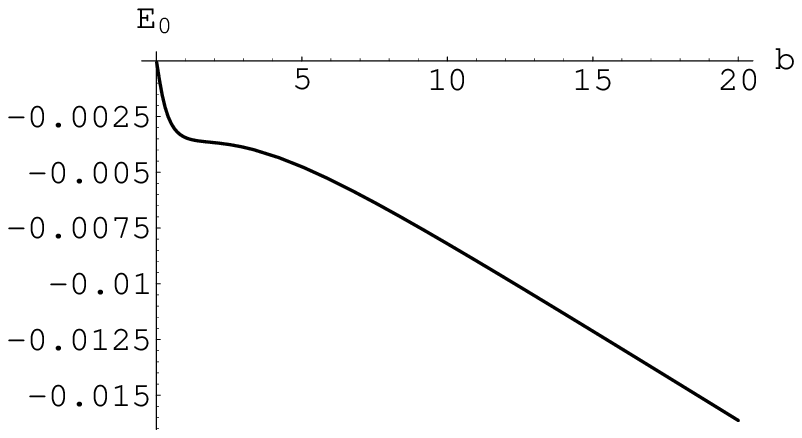}}
\caption{{\protect\small In the fermionic case, plot of the energy in
terms of the dimensionless  magnetic field $b$, for a fixed value
of the dimensionless mass $\mu$ (here $\mu =1$). In Fig. 4a, detail
the formation of an  inflexion region, and the
inflection point; in Fig. 4b, the intermediate region; finally, in Fig. 4c,
asymptotic behavior for large values of $b$.}} \label{f4}
\end{figure}
\begin{figure}[h]
\centerline{\epsfxsize=8cm \epsfbox{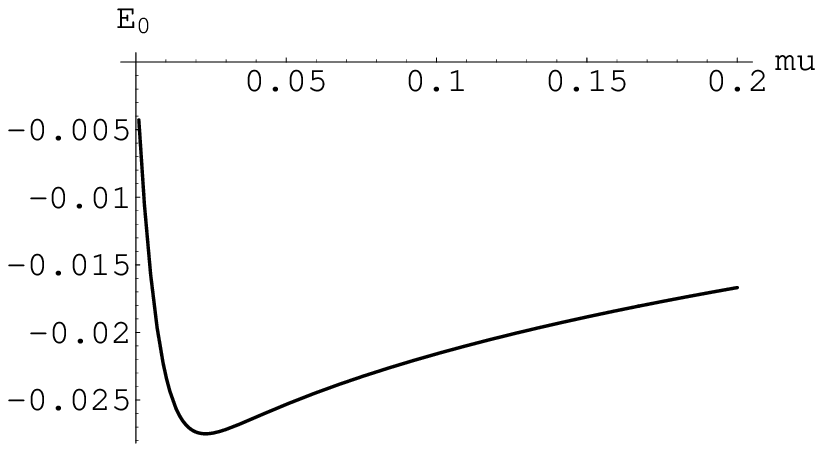} \hspace{8mm}
\epsfxsize=8cm \epsfbox{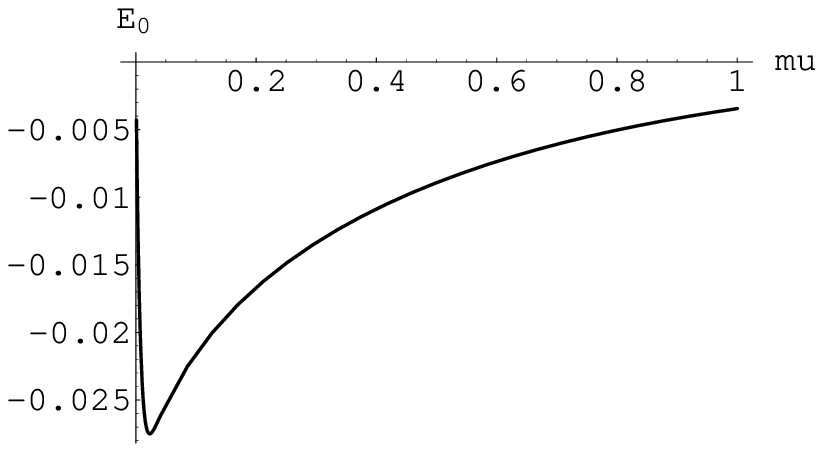}}
\caption{{\protect\small Plot of the energy in terms of the dimensionless
mass $\mu$. In Fig. 5a we see the behavior for small mass, with the 
formation of a smooth minimum, which looks much sharper in Fig. 5b, where the
asymptotic behavior for large $\mu$ is clearly established.}} \label{f5}
\end{figure}

The result of the numerical evaluation of Eq. (\ref{NumAnalF}) for
a sample of values of \( \mu  \) and \( b \) is shown in Figs. 4
and 5. Fig. 4 shows the dependence of the energy on the magnetic field
\( b \) for different ranges of this quantity, to better show the
inflexion region (Fig. 4a, low values of \( \mu  \)), the intermediate
region (Fig. 4b) and the asymptotic behavior for large values of \( b \)
(Fig. 4c). The intermediate region is most interesting, showing a
smooth transition from the case of small magnetic field (with a good
zero field limit) to the linear asymptotic behavior corresponding
to a large field. And this situation is common for any particular
value of the mass \( \mu  \). The transition is given by a smooth
inflexion point of nearly horizontal tangent. It looks like a minimum
could almost be formed, but not quite. In Fig. 5 the dependence of
the energy on the dimensionless mass \( \mu  \) is depicted, showing
the behavior in the region of small mass (Fig. 5a) and the asymptotic
behavior for large \( \mu  \) (Fig. 5b). Needless to say, the asymptotic
behavior corresponds to the the analytic expressions obtained before.
For the intermediate region we see a smooth behavior connecting the
two regions with a minimum that is very easily obtained with good
precision, for any particular value of \( b \).

The same observations as for the bosonic case can be done here, namely,
that the graphs shown are only a very small sample of the ones that
can be easily obtained from our general formulas in this section,
corresponding to the fermionic case, for different values of the variables,
and including two-dimensional plots. We should warn the reader, however,
that the computation time is drastically increased now.

\section{Final remarks}

In this paper we have considered the influence of an external uniform
magnetic field on the Casimir energy associated with charged quantum
fields confined by hard boundary conditions. The method employed incorporates,
in a relatively simple way, the hard boundary conditions of the Dirichlet
or MIT type imposed, respectively, on a massive bosonic field and a
massive fermionic field plus, for both cases, the effect of the magnetic
field, and leads to expressions for the vacuum energy especially suited
for numerical calculations. The analytical and numerical results show
that, in the bosonic case and Dirichlet boundary conditions, the effect
of a large, applied magnetic field is to suppress the vacuum energy,
and in the case of fermionic field confined in a slab-bag with MIT
boundary conditions, the effect of the applied field is to enhance
(in absolute magnitude) the vacuum energy. For small and for intermediate
values of the scaled magnetic field, however, a more interesting behavior
of the vacuum energy shows up. Since here the calculations were performed
in the general framework of an effective field theory, the details
of the interaction between the quantum field and the external
magnetic field ---represented by the generation of the Landau levels
and the hard boundary conditions--- remains hidden. It would be interesting
to understand the behavior of the vacuum energy under the circumstances
considered here in terms of a more fundamental model of the structure
of the confined quantum vacuum.

\section*{Acknowledgments}

A.C.T. wishes to acknowledge the hospitality of the Institut d'Estudis
Espacials de Catalunya (IEEC/CSIC) and the Universitat de Barcelona,
Departament d'Estructura i Constituents de la Mat\`{e}ria and the
financial support of CAPES, the Brazilian agency for faculty improvement,
Grant Bex 0168/01-2. The investigation of E.E. has been supported
by DGI/SGPI (Spain), project BFM2000-0810, and by CIRIT (Generalitat
de Catalunya), contract 1999SGR-00257.

\end{document}